\pdfoutput=1
\documentclass[sigconf,nonacm]{acmart}

\renewcommand\footnotetextcopyrightpermission[1]{}
\usepackage{booktabs}
\usepackage{balance}

\begin{document}

\title{Open Source Is Not One Thing: A Typology of Open-Source Software Sub-Genres}

\author{Mohamed Ouf}
\email{24blr2@queensu.ca}
\affiliation{
  \institution{Queen's University}
  \city{Kingston}
  \state{Ontario}
  \country{Canada}}

\author{Rowan Hussein}
\email{rhuss060@uottawa.ca}
\affiliation{
  \institution{University of Ottawa}
  \city{Ottawa}
  \state{Ontario}
  \country{Canada}}

\renewcommand{\shortauthors}{Ouf and Hussein}

\begin{abstract}
Open source software (OSS) is not homogeneous. A project's purpose, governance, and
funding shape how its community forms, who contributes, and how the software is
maintained, yet empirical research often samples OSS broadly and reports findings as if
they held for open source as a whole. We argue that OSS comprises distinguishable
sub-genres, and that the sub-genre a study samples bounds how far its findings generalize.
Using a light, multi-source review that screens 3{,}925 unique papers, we synthesize a
typology of fourteen OSS sub-genres, from well-studied ones such as community-driven,
company-backed, foundation-governed, research and scientific, and open source for social
good (OSS4SG), to under-studied ones such as multi-company co-opetition, protestware, and
open-source appropriate technology. We place the sub-genres in a framework that records each one's
primary driver, governance, and funding, with its maturity in the literature and
representative projects, and we present a research agenda whose central question is whether findings established on one
sub-genre transfer to others. The contribution is the typology and the agenda rather than a
complete census, and we mark the sub-genres whose empirical support is thin.
\end{abstract}



\maketitle

\section{Introduction}
\label{sec:intro}

Open source software (OSS) is not one thing. Empirical research nonetheless samples it as
though it were. Studies draw a set of popular GitHub projects, fit models of contributor
retention, onboarding, or code quality, and report the results for open source in general.
The implicit assumption is that a community Linux distribution, a single-vendor database, a
university research library, a humanitarian health-records system, and a one-person utility
differ only in size and popularity. They do not. They differ in why the project exists, who
decides its direction, how the work is paid for, and how the community behaves.

Recent evidence makes the difference concrete. A comparison of mission-driven open source
for social good (OSS4SG) and conventional OSS reports different community structure,
contributor retention, and code-quality management, with OSS4SG communities retaining
contributors (``sticky'') and conventional projects attracting many who do not stay
(``magnetic'') \cite{ouf2026community}. The newcomer-to-core pathway also differs by
sub-genre, and contributors in OSS4SG reach core status at higher rates and through more
than one route \cite{ouf2026doGood}. Even the way a newcomer first joins, whether through a
mentorship program, a hackathon, or independently, is associated with how long they stay
\cite{ouf2026sameStart}. When two sub-genres diverge this far, a retention model or an
onboarding program validated on one of them need not hold for the others. Coarse metrics
such as stars and commit counts hide the structure that decides whether a finding
generalizes.

The distinction matters in practice as well. A developer deciding where to contribute
gains from knowing what kind of project they are approaching, because its governance,
funding, and community behavior set what a first contribution takes and how a newcomer is
received. Social barriers at the first contribution already push newcomers to abandon
projects \cite{steinmacher2015barriers}, and the path into a single-vendor product differs
from the path into a volunteer distribution or a research library. Sub-genre bears on sustainability too. How a project is funded and
who maintains it, and so how resilient it is, ranges from firm-employed teams to the few
unpaid volunteers behind widely used infrastructure \cite{eghbal2016roads,coelho2017fail}.
Naming a project's sub-genre makes these differences explicit instead of leaving each
contributor and each study to rediscover them.

This paper makes three contributions. First, we present a typology of fourteen OSS
sub-genres (Section~\ref{sec:subgenres}), organized by who drives a project and to what
end. Second, we place the sub-genres in a framework that records their driver, governance, and
funding (Section~\ref{sec:framework}). Third, we present a research agenda that treats sub-genre as
a variable to report and control for in empirical OSS studies (Section~\ref{sec:agenda}).
The paper is short by design. We mark the sub-genres whose evidence is thin as targets for
future work.

\section{Related Work}
\label{sec:related}

Prior work classifies open source along single axes. \citet{fitzgerald2006transformation}
contrasts a volunteer ``OSS 1.0'' with a commercially co-opted ``OSS 2.0''.
\citet{riehle2012singlevendor} separates community open source from single-vendor
commercial open source, and \citet{omahony2007governance} contrasts community-managed
projects with firm-involved hybrids. Other schemes classify by the firm-community
relationship \cite{dahlander2005relationships}, by governance configuration
\cite{ditullio2013governance,markus2007governance}, by business model
\cite{duparc2022archetypes}, by contributor motivation \cite{shah2006motivation}, or by
structural growth dynamics, as in the stadiums, clubs, federations, and toys of
\citet{eghbal2020working}. A broad survey of free and open-source development covers the
literature across topics from individual contributions to governance and community dynamics \cite{crowston2012freelibre}.

These schemes are valuable but partial. Each cuts the space on one dimension, and most
resolve into a binary or a few types that center on commercial form and leave little room
for non-commercial and public-interest projects. The typology presented here differs in two ways. It
organizes OSS by who drives a project and to what end, an axis that the
commercial-versus-community binaries approximate but do not fully separate, and it spans
fourteen sub-genres that include forms the earlier schemes collapse or omit, such as
multi-company co-opetition, government and civic OSS, open-source appropriate technology,
protestware, and critical digital infrastructure.

\section{Method}
\label{sec:method}

We conducted a light, multi-source review rather than an exhaustive systematic one, and we
release the search-and-screen script so the corpus can be regenerated. We queried two
scholarly indexes, OpenAlex and arXiv, with 46 queries grouped into fifteen facets, one
facet per candidate sub-genre plus a discovery facet aimed at sub-genres we had not
anticipated, such as ``taxonomy of open source projects''. The search returned 5{,}771
records. We normalized them to a common schema and removed duplicates by DOI or title,
which left 3{,}925 unique papers. We then screened the papers with a rule that keeps a
paper when its title or abstract both refers to open source and signals that the work
defines, classifies, or studies a kind of OSS project or community. Of these, 399 papers
passed. Coverage across the fourteen sub-genres ranged from 8 papers, for critical digital
infrastructure, to 76, for OSS4SG, and reflects query reach rather than the size of each
field. We grouped the screened papers into the sub-genres below. The search-and-screen
design follows the bibliometric and systematic-review practice used in prior work
\cite{ouf2024maritimeDT,ouf2024ris}.

We treat the resulting typology as a working hypothesis to be confirmed, split, merged, and
extended, not as a settled ontology, and Section~\ref{sec:threats} states its limits.

\section{A Typology of OSS Sub-Genres}
\label{sec:subgenres}

\begin{table*}[t]
\centering
\caption{A typology of OSS sub-genres, recording each one's primary driver, governance, funding, maturity in the literature, and representative projects. Maturity reflects the depth of
dedicated empirical literature, where established marks a substantial body of work, growing
marks an active but smaller one, and emerging marks only a handful of dedicated studies. The
sub-genres are not mutually exclusive, and many real projects span more than one.}
\label{tab:typology}
\footnotesize
\begin{tabular}{@{}p{2.6cm}p{2.7cm}p{2.5cm}p{2.4cm}p{1.4cm}p{2.5cm}@{}}
\toprule
\textbf{Sub-genre} & \textbf{Primary driver} & \textbf{Governance} & \textbf{Funding} &
\textbf{Maturity} & \textbf{Examples} \\
\midrule
Community-driven & Intrinsic: learning, reputation, ideology & Meritocratic, peer-elected & Volunteer, donations & Established & Debian, Arch Linux, GNU Emacs \\
Company-backed OSS & Commercial: revenue, market control & One firm controls & Firm-employed staff, product revenue & Established & Elastic, MongoDB, GitLab \\
Foundation-governed & Neutral multi-party collaboration & Foundation rules, vendor-neutral & Member dues, corporate sponsorship & Established & Apache HTTP, Kubernetes, Eclipse \\
Multi-company co-opetition & Strategic shared work among rivals & Shared across firms & Competing firms' staff & Emerging & OpenStack, Linux kernel \\
InnerSource & Internal reuse, efficiency & Trusted committers within a firm & Corporate, internal & Growing & HP, Philips, Ericsson \\
OSS for Social Good (OSS4SG) & Societal benefit & NGO, nonprofit, community & Grants, donations, NGOs & Growing & OpenMRS, DHIS2, Ushahidi \\
Government and civic & Public mandate, transparency & Publicly accountable & Public funds, procurement & Growing & Decidim, X-Road, GovStack \\
Educational & Learning and social good & Faculty or instructor led & Universities, course staff & Growing & Sahana Eden, LibreFoodPantry \\
Open-source appropriate technology & Basic needs, sustainability & Practitioner and academic networks & Grants, NGOs & Emerging & RepRap, WikiHouse \\
Hobbyist and solo & Personal enjoyment, learning & None, single maintainer & Unpaid, occasional donations & Established & left-pad, micrograd \\
Protestware & Political or economic protest & Maintainer acts unilaterally & Unpaid & Emerging & node-ipc, colors.js \\
Research and scientific (RSE) & Research support, reproducibility & Laboratory or PI led & Research grants & Established & Astropy, scikit-learn, LAMMPS \\
Open hardware and open data & Reproducibility, maker, civic data & Domain consortia, communities & Grants, community & Growing & Arduino, OpenStreetMap \\
Critical digital infrastructure & Public good, technical interest & Informal, few maintainers & Under-funded, ad-hoc sponsorship & Growing & OpenSSL, curl, log4j \\
\bottomrule
\end{tabular}
\end{table*}

We organize the sub-genres by their primary driver, that is, who starts and sustains the
project and to what end. The axis is not perfectly clean, since many projects span more
than one sub-genre, but it captures the main source of variation.
Table~\ref{tab:typology} summarizes all fourteen sub-genres by primary driver, governance, and funding, with maturity and examples. The
subsections below define each sub-genre and cite representative work, and we mark the ones
whose dedicated literature is still thin as emerging.

\subsection{Firm-driven open source}

\textbf{Company-backed OSS} is owned and primarily developed by one firm that earns revenue
from the software, through an open-core model (a free core paired with paid proprietary
extensions), dual licensing, or a hosted service, while the firm controls the project's
direction. \citet{riehle2012singlevendor} defines the single-vendor commercial form, and
later studies analyze how firms participate in and govern such projects
\cite{west2008participation,dahlander2008firms}. \textbf{Multi-company co-opetition} is an
emerging sub-genre in which firms that compete in the market jointly develop shared
infrastructure. \citet{teixeira2016cooperation} examined cooperation among competing firms
in OpenStack \cite{teixeira2015lessons}, and recent work characterizes how such
collaborations are governed \cite{osborne2025characterising}. \textbf{InnerSource} applies
open-source practices inside a single organization, adding open contribution and
trusted-committer review to internal development for cross-team reuse
\cite{capraro2016innersource,stol2014keyfactors,stol2015tutorial}.

\subsection{Community- and foundation-governed open source}

\textbf{Community-driven OSS} is the classic peer-production model, built by volunteers who
contribute from intrinsic motivation and where authority follows merit rather than
ownership. \citet{benkler2002coase} established the economic theory of commons-based peer
production, and empirical studies describe the social and authority structure of projects
such as Debian \cite{crowston2005social,omahony2007emergence}. \textbf{Foundation-governed
OSS} places a project under a vendor-neutral nonprofit, such as the Apache Software
Foundation, the Linux Foundation, or the Eclipse Foundation, that holds the trademark and
intellectual property and enforces neutral rules so that competing firms can collaborate
\cite{canovas2018role,duenas2007apache,yang2022apache}.

\subsection{Mission- and impact-driven open source}

\textbf{Open source for social good (OSS4SG)} comprises projects whose primary purpose is
societal benefit rather than commercial value, such as health-records systems and
crisis-response tools, and many appear in digital public goods registries.
\citet{huang2021fingerprints} studies the motivations and challenges of contributing to
OSS4SG, and later work examines its contributors and community dynamics
\cite{fang2023fouryear,ouf2026community,ouf2026doGood}. \textbf{Government and civic OSS},
including software designated as a digital public good, is commissioned or mandated by
public institutions under transparency and procurement constraints
\cite{linaker2023publicsector25years,linaker2025pubsectordev,vanloon2015adopting}.
\textbf{Educational OSS} is produced inside teaching contexts, with students contributing
for course credit. The Humanitarian Free and Open Source Software (HFOSS) tradition is its
archetype, pairing software-engineering education with socially beneficial projects
\cite{ellis2007humanitarian,hislop2015multiinstitutional,pinto2019training}.
\textbf{Open-source appropriate technology (OSAT)} is an emerging sub-genre that extends
open-source practice from software to hardware and design knowledge meant to meet basic
needs in resource-limited settings. \citet{pearce2012case} articulates OSAT and
demonstrates it in practice \cite{pearce2010printing,pearce2012building}.

\subsection{Individual- and conflict-driven open source}

\textbf{Hobbyist and solo OSS} consists of projects that one person builds for enjoyment or
to meet a personal need, with no formal governance and a low truck factor, meaning that few
contributors would have to leave for the project to stall
\cite{kalliamvakou2014promises,avelino2016truck,coelho2017fail}. \textbf{Protestware} is an
emerging sub-genre in which a maintainer alters or sabotages a package to deliver a
political or economic message, so that the defining act is conflict rather than feature work
\cite{fan2025protestware,fan2024going,cheong2023ethical}.

\subsection{Knowledge- and infrastructure-driven open source}

\textbf{Research and scientific software (RSE)} is built inside research to support and
reproduce studies, usually by domain scientists rather than trained software engineers, and
it is funded by grants. It is at once a research output and a research instrument
\cite{wilson2014bestpractices,hannay2009scientists,heaton2015claims}. \textbf{Open-source
hardware and open data} apply open-source practice to non-software artifacts, including
physical designs and public datasets
\cite{bonvoisin2018participative,boujut2019communities,budhathoki2013motivation}. Some
projects release an open dataset as the primary artifact, such as in
computer-vision-based fitness analysis \cite{ouf2024visionpf}. \textbf{Critical digital
infrastructure OSS} comprises components that much of the software ecosystem depends on
indirectly, such as OpenSSL and curl, yet that a few volunteers often maintain.
\citet{eghbal2016roads} named this sustainability problem, and later work formalizes it as
underproduction, where the labor a component receives falls short of how widely it is
relied upon \cite{champion2021underproduction}. Studies of the npm ecosystem also show how
a few maintainers' accounts reach much of it \cite{zimmermann2019smallworld}.

\section{A Typology Framework}
\label{sec:framework}

Table~\ref{tab:typology} shows two patterns. First, the primary-driver axis does not fix
the other dimensions. Foundation-governed OSS and multi-company co-opetition both involve
competing firms, yet they differ in where authority sits. Second, the maturity column shows
where evidence is thin. Several sub-genres that matter in practice, namely multi-company
co-opetition, protestware, and open-source appropriate technology, rest on a few studies
each, against the deep literatures on community-driven, company-backed, and
foundation-governed OSS and on research and scientific software.

\section{Why It Matters: A Research Agenda}
\label{sec:agenda}

\paragraph{Transfer of findings across sub-genres.}
This is the question the typology raises. A retention model built on company-backed OSS, an
onboarding program validated on a large community project, or a code-quality heuristic drawn
from popular repositories need not transfer to research software, humanitarian projects, or
solo-maintained infrastructure. Prior work already shows that contributor retention, the
newcomer-to-core pathway, and the effect of entry events differ between OSS4SG and
conventional OSS \cite{ouf2026community,ouf2026doGood,ouf2026sameStart}. How far this
difference extends across the other sub-genres is open. Research and scientific software is
a concrete next case, since its contributors work to research calendars and academic
careers rather than product cycles, and whether community-structure findings from
conventional OSS hold there is untested.

\paragraph{Under-studied sub-genres.}
Critical digital infrastructure is the clearest case, a sub-genre the whole ecosystem
depends on yet one defined by few maintainers and limited funding
\cite{eghbal2016roads,champion2021underproduction}. Multi-company co-opetition and
protestware are similarly consequential and similarly thin. A sub-genre-aware research
program would direct effort toward these cases.

\paragraph{Reporting and controlling for sub-genre.}
A study should report the sub-genre it samples, and control for it where possible, as
studies already report programming language or project size. Tooling can help. Methods that
recover intent from artifacts, such as generating user stories from source code with large
language models \cite{ouf2025userStories}, point toward classifying projects into
sub-genres at scale, which would make sub-genre-stratified analysis routine. Classification
at scale would also serve practice, letting a contributor see what kind of community they
are joining before investing effort, and letting maintainers and funders find the
sub-genres where support is thin.

\section{Threats to Validity}
\label{sec:threats}

The review is light rather than systematic. It covers English-language and well-indexed
venues, so we may have missed sub-genres visible mainly in other languages or in grey literature. The sub-genres are not mutually exclusive. Kubernetes is at once
foundation-governed and multi-company co-opetitive, and OpenMRS is at once OSS4SG and
educational, so the sub-genres overlap rather than partition the space. The maturity labels
reflect the literature we retrieved rather than a citation count, and they can understate a
sub-genre whose evidence is recent or scattered. The per-sub-genre screened counts reflect
the reach of the queries, not the true size of each field. The dimensions are deliberately
coarse, and we leave finer stratification, such as by application domain, to future work.

\section{Conclusion}
\label{sec:conclusion}

Open source is plural. Treating it as one thing lets coarse metrics hide differences in
purpose, governance, funding, and community behavior, differences large enough that a
finding from one sub-genre need not hold for another. We have presented a typology of
fourteen OSS sub-genres, placed them in a comparison framework, and proposed a research
agenda whose first task is to test how far existing findings generalize. The typology is
provisional. Its value is in making the variety of open source explicit, and in marking the
emerging sub-genres, namely co-opetition, protestware, and appropriate technology, where new
empirical work is most needed.

\balance
\bibliographystyle{ACM-Reference-Format}
\bibliography{references}

\end{document}